\documentclass[twocolumn,aps,prl,superscriptaddress,amsfonts]{revtex4-1}
\usepackage{amsmath}
\usepackage{graphicx}
\usepackage{bm}
\usepackage{array}
\usepackage{dcolumn}
\newcommand{\ket}[1]{{| {#1} \rangle}}

\begin{document}

\title{Observation of Prethermalization in Long-Range Interacting Spin Chains}

\author{B. Neyenhuis}
\altaffiliation[Present address: ]
{Lockheed Martin Corporation,  Littleton, CO 80127, USA}
\email{brian.neyenhuis@lmco.com}
\affiliation{Joint Quantum Institute and Joint Center for Quantum Information and Computer Science, University of Maryland Department of Physics and National Institute of Standards and Technology,
College Park, Maryland 20742}
\author{J. Smith}
\affiliation{Joint Quantum Institute and Joint Center for Quantum Information and Computer Science, University of Maryland Department of Physics and National Institute of Standards and Technology,
College Park, Maryland 20742}
\author{A. C. Lee}
\affiliation{Joint Quantum Institute and Joint Center for Quantum Information and Computer Science, University of Maryland Department of Physics and National Institute of Standards and Technology,
College Park, Maryland 20742}
\author{J. Zhang}
\affiliation{Joint Quantum Institute and Joint Center for Quantum Information and Computer Science, University of Maryland Department of Physics and National Institute of Standards and Technology,
College Park, Maryland 20742}
\author{P. Richerme}
\affiliation{Department of Physics, Indiana University, Bloomington, IN, 47405}
\author{P. W. Hess}
\affiliation{Joint Quantum Institute and Joint Center for Quantum Information and Computer Science, University of Maryland Department of Physics and National Institute of Standards and Technology,
College Park, Maryland 20742}
\author{Z.-X. Gong}
\author{A. V. Gorshkov}
\author{C. Monroe}
\affiliation{Joint Quantum Institute and Joint Center for Quantum Information and Computer Science, University of Maryland Department of Physics and National Institute of Standards and Technology,
College Park, Maryland 20742}

\date{\today}

\begin{abstract}
Statistical mechanics can predict thermal equilibrium states for most classical systems, but for an isolated quantum system there is no general understanding on how equilibrium states dynamically emerge from the microscopic Hamiltonian \cite{Polkovnikov2011,Cazalilla2010,Kinoshita2006,Manmana2007,Babadi2015,Anderson1958,Basko2007,Schreiber2015,Smith2016}. For instance, quantum systems that are near-integrable usually fail to thermalize in an experimentally realistic time scale and, instead, relax to quasi-stationary prethermal states that can be described by statistical mechanics when approximately conserved quantities are appropriately included in a generalized Gibbs ensemble (GGE) \cite{Gring2012,Langen2015,Rigol2007,Ilievski2015}. Here we experimentally study the relaxation dynamics of a chain of up to 22 spins evolving under a long-range transverse field Ising Hamiltonian following a sudden quench. For sufficiently long-ranged interactions the system relaxes to a new type of prethermal state that retains a strong memory of the initial conditions. In this case, the prethermal state cannot be described by a GGE, but rather arises from an emergent double-well potential felt by the spin excitations. This result shows that prethermalization occurs in a significantly broader context than previously thought, and reveals new challenges for a generic understanding of the thermalization of quantum systems, particularly in the presence of long-range interactions \cite{Gong2013}.
\end{abstract}

\maketitle

Statistical mechanics can predict thermal equilibrium states for most classical systems, but for an isolated quantum system there is no general understanding on how equilibrium states dynamically emerge from the microscopic Hamiltonian. For instance, quantum systems that are near-integrable usually fail to thermalize in an experimentally realistic time scale and instead, relax to quasi-stationary prethermal states that can be described by statistical mechanics when approximately conserved quantities are appropriately included in a generalized Gibbs ensemble (GGE). Here we experimentally study the relaxation dynamics of a chain of up to 22 spins evolving under a long-range transverse field Ising Hamiltonian following a sudden quench. 

In the classical world thermalization is expected in all but special cases where conserved quantities or hidden symmetries prevent the ergodic exploration of phase space. Because the classical world is ultimately comprised of quantum systems we therefore expect that a closed quantum system will also reach thermal equilibrium. Although quantum dynamics are unitary, measurements made within a subsystem trace over the rest of the system and appear thermal because the rest of the system acts as a thermal bath \cite{vandenWorm2013, Reimann2008, Kaufman2016,Neill2016}.  


However this is not always the case. For integrable models an extensive number of conserved quantities prevent the efficient exploration of phase space \cite{Kinoshita2006, Manmana2007} and the system relaxes to a steady-state predicted by a GGE \cite{Rigol2007, Ilievski2015} specified by the initial values of the integrals of motion. For near-integrable systems, such as weakly interacting ultracold gases, thermalization can still occur, but only over extremely long time scales beyond current experimental reach \cite{Gring2012,Langen2015}. However, it is possible to observe  quasi-stationary states, often called prethermal, that emerge within an experimentally accessible time scale. 

\begin{figure}
\includegraphics*[width=8cm]{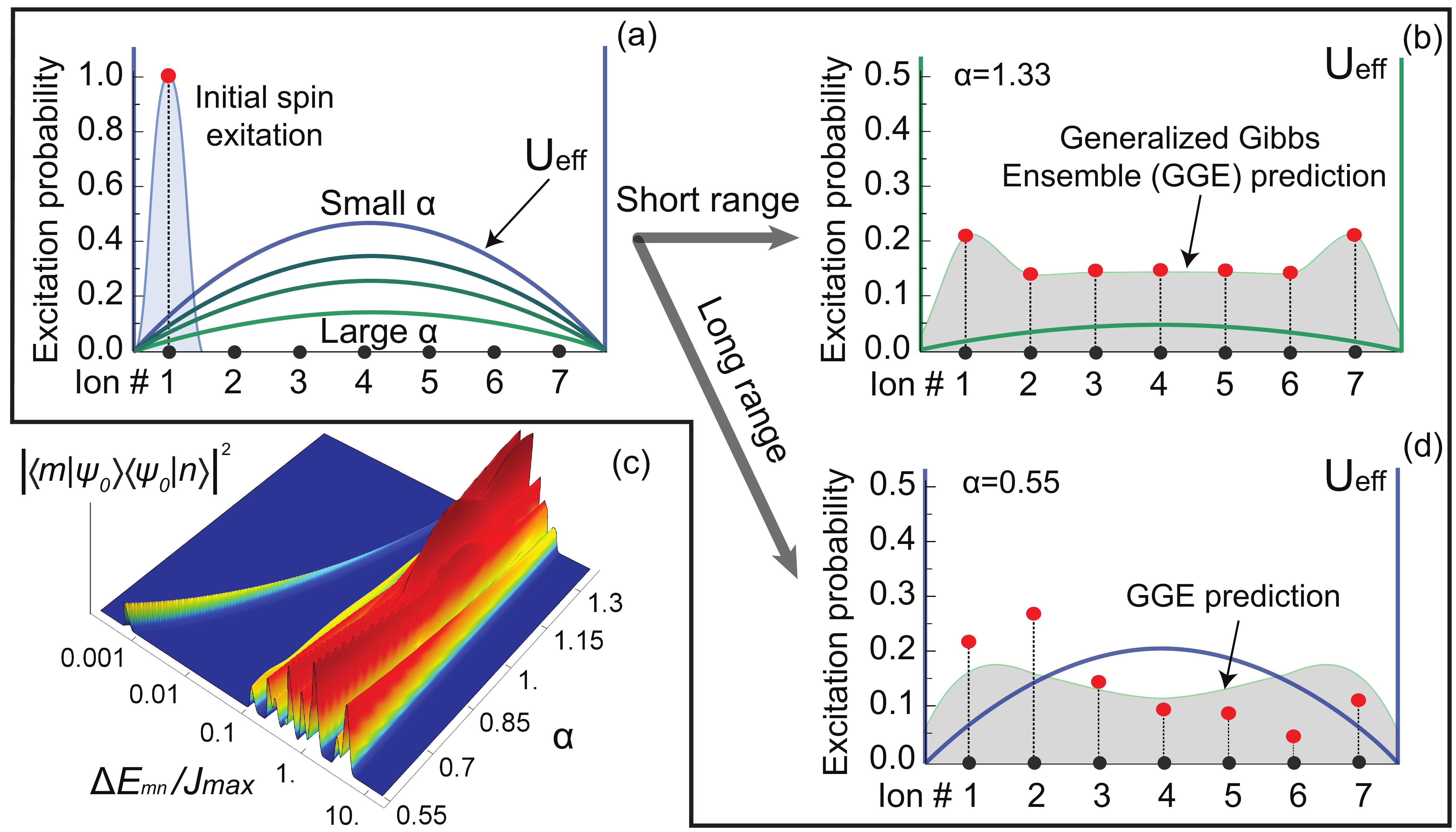}
\caption{\textbf{Emergent double well potential}. (a) The spin chain starts with a single spin excitation on the left end in an effective double-well potential, Ueff, whose barrier height is determined by the range $\alpha$ of the interactions. (b) For short range interactions ($\alpha = 1.33$), we map the system to a particle in a 1D square well where the excitation becomes symmetrically distributed across the chain as predicted by the GGE, $\langle \sigma_i^z\rangle_{GGE}$. (d) However, for long-range interactions ($\alpha = 0.55$) there is an emergent double-well potential which prevents the efficient transfer of the spin, and the excitation location retains memory of the initial state, in contrast to $\langle \sigma_i^z\rangle_{GGE}$. (c) The double well gives rise to near-degenerate eigenstates as $\alpha$ is decreased as seen in the calculated energy difference between all pairs of eigenstates versus $\alpha$, with amplitude weighted by the product of the eigenstates' overlap with the initial state.}
\label{fig1main}
\end{figure}

Previous observations of prethermal states have focused on those described by a GGE associated with the integrable part of the model \cite{Gring2012,Langen2015}. Here, we observe a new form of prethermalization \cite{Berges2004}, where a system of interacting spins rapidly evolves to a quasi-stationary state that cannot be predicted by a GGE. This type of prethermal state arises, even in the thermodynamic limit, when a system has long-range interactions and open boundaries such that the translational invariance is broken. As a result, spin excitations feel an emergent double-well potential whose depth grows with interaction range (Fig.\ref{fig1main}). Memory of the initial state is preserved by this emergent potential, but is eventually lost due to weak tunneling between the two wells and the interactions between spin excitations.



Effective spin-1/2 particles are encoded in the $^2$S$_{1/2}|F=0,m_F=0\rangle$ and $|F=1,m_F=0\rangle$ hyperfine `clock' states of a $^{171}$Yb$^{+}$ ion, denoted $\lvert\downarrow\rangle_{z}$ and $\lvert\uparrow\rangle_{z}$ \cite{Olmschenk2007}.  We confine a chain of ions in a linear rf Paul trap and apply optical dipole forces to generate the effective spin-spin coupling \cite{Molmer1999,Kim2009} of an Ising Hamiltonian (methods):
\begin{equation}
H=   \sum_{i <  j} J_{i j} \sigma^x_i \sigma^x_j + B \sum_i \sigma^z_i, 
\label{eq:Ham}
\end{equation}
where $\sigma_i^\gamma$ ($\gamma=x,z$) are the Pauli matrices acting on the $i^\text{th}$ spin, $J_{ij}$ is the coupling between spins $i$ and $j$, $B/(2 \pi)=10$ kHz is a uniform effective transverse field, and we use units in which Planck’s constant equals 1 (methods). The spin-spin interaction is long-range and can be described by a power law decay where $J_{i j} = \frac{J_{\textrm{max}}}{\vert i-j\vert^\alpha}$,  $J_{\textrm{max}}/(2 \pi)$ is the maximum coupling strength which ranges from 0.45 kHz to 0.98 kHz. We tune the power law exponent $\alpha$ between $ 0.55 $ and $ 1.33$ by changing the axial confinement of the ions. With long-range interactions, $H$ is in general non-integrable (in contrast to the nearest-neighbor case where the 1D model is integrable \cite{Ising1925}) and thermalization is anticipated in the long time limit according to the eigenstate thermalization hypothesis \cite{Srednicki1994,Rigol2008,Palma2015}. However, one can map (\ref{eq:Ham}) with a small number of spin excitations into a near-integrable model of bosons, with an integrable part made of free bosons and an integrability-breaking part consist of weak interactions among bosons that leads to the eventual thermalization.

We initialize the chain of seven spins by optically pumping all spins to the $\lvert\downarrow\rangle_{z}$ state, and then use a tightly focused individual-ion addressing laser to excite a single spin on one end of the chain to the $\lvert\uparrow\rangle_{z}$ state as seen in Fig. \ref{fig1main}a (methods) \cite{Lee2016}. The spins then evolve under (\ref{eq:Ham}) and we measure the time evolution of the spin projection in the $z$-basis. For the shortest range interactions we realize ($\alpha = 1.33$) the system rapidly evolves to a prethermal state predicted by the GGE associated with the integrals of motion corresponding to the momentum space occupation number of the non-interacting bosons, which does not preserve memory of the initial spin excitation location (methods) (Fig 1b).

However, in the long-range interacting case ($\alpha=0.55$) we see the position of the spin excitation reaches an equilibrium value that retains a memory of the initial state (Fig. \ref{fig1main}d) out to the longest experimentally achievable time of $25/J_{\textrm{max}}$. This prethermal state is in obvious disagreement with both a thermal state and the GGE prediction, which both maintain the right-left symmetry of the system. 


The dynamics of the spin-wave boson model for short-range interactions are similar to those of a free-particle in a square-well potential. However, long-range interactions distort the square-well to a double-well potential Fig. \ref{fig1main}a. Here we emphasize that the double-well potential emerges non-trivially because our model (\ref{eq:Ham}) is transitionally invariant without boundaries. The spin-wave boson model has an extensive number of near-degenerate eigenstates that are symmetric and antisymmetric superpositions of spin excitations in the left and right potential wells. For seven spins, we calculate the energy difference, $\Delta E_{\text{mn}}$, between all pairs of eigenstates and plot them with respect to $\alpha$ in Fig.\ref{fig1main}c where the amplitudes, $|\langle m|\psi_{0}\rangle \langle\psi_{0}|n\rangle|^{2}$, are products of the overlaps of the eigenstates $|m\rangle$ and $|n\rangle$ overlap with the initial state$|\psi_{0}\rangle$.  With $\alpha=0.55$ the two lowest energy states are almost degenerate, with an energy difference of approximately one-thousand times smaller than $J_{\text{max}}$, due to the tunneling rate between the ground states of each well, which is exponentially small in the barrier height. As a result the spin excitation will remain in its initial well until it tunnels across the potential barrier at much longer times. 



\begin{figure*}
\includegraphics[width=\textwidth]{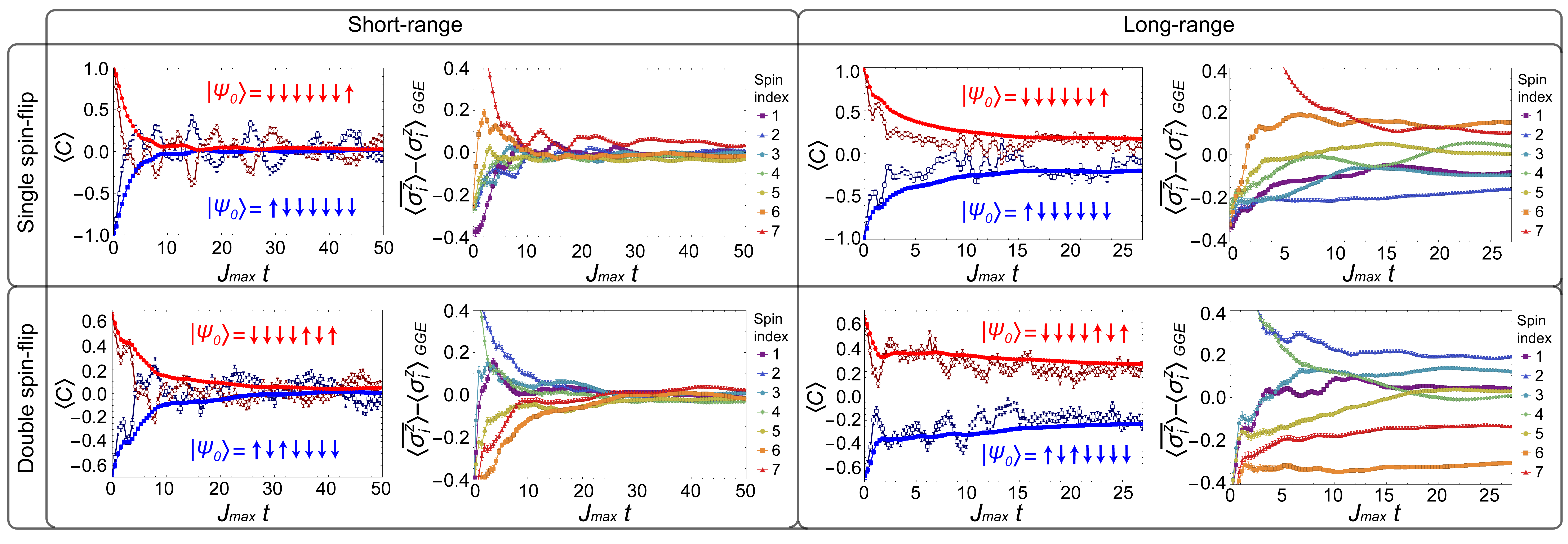}
\caption{\textbf{Measured location of spin excitation}. The average position of the spin excitation, $\langle C \rangle$, is plotted for an initial excitation on the left (dark blue unfilled squares) and right (dark red unfilled circles) along with their cumulative time average (blue and red filled circles and squares), for short-range ($\alpha = 1.33$) and long-range ($\alpha = 0.55$) interactions with initial states with one spin excitation (top panels) and two spin excitations (bottom panels). The cumulative time average of the deviation of the postselected individual spin projections from the generalized Gibbs ensemble, $\langle \overline{\sigma_i^z}\rangle - \langle \sigma_i^z \rangle_{GGE}$, is also plotted. For short-range interactions the spins quickly thermalize to a value predicted by the GGE, but for the long-range interacting system a new type of prethermal state emerges.  Error bars, 1 s.d.}
\label{fig2main}
\end{figure*}

To better characterize the ``location'' of the spin excitation we construct the observable
\begin{equation}
C = \sum_i \frac{2 i-N-1}{N-1} \frac{\sigma_i^z+1}{2},
\end{equation} 
where $N$ is the number of ions. The expectation value of $C$ varies between -1 and 1 for a spin excitation on the left and right ends, respectively. Due to the spatial inversion symmetry of the spin-wave boson model and (\ref{eq:Ham}), both the GGE predicted and thermal values of $\langle C\rangle$ are zero. In Fig. \ref{fig2main} we plot $\langle C \rangle$ along with its cumulative time average $\langle\overline{C}\rangle$ for $\alpha=1.33$ (short-range) and $\alpha = 0.55$ (long-range).  To further accentuate the asymmetry of the prethermalization we prepare initial states with a single-spin excitation on the left or right ends of the chain. 

With short-range interactions the system quickly relaxes to the prethermal value predicted by the GGE where $\langle C \rangle$ is near zero irrespective of the initial state.  But with long-range interactions, the prethermal state that emerges retains a clear memory of the initial conditions and is different than both the thermal and GGE predictions.

It is useful to talk about thermalization in terms of local quantities since subsystems must use the rest of the chain as a heat bath. In Fig. \ref{fig2main} we plot the cumulative time average of the deviation of the single spin magnetizations, $\langle \sigma_{i}^{z} \rangle$, from the equilibrium value predicted by the GGE. Here we postselect the data for the correct number of spin excitations to eliminate the effects of imperfect state preparation and small deviations from our model Hamiltonian due to unwanted excitations of the phonon modes (methods).  For the short-range interactions we see the cumulative time average for each spin quickly converge to the GGE predicted value. However, with the long-range interactions we see the individual spins reach a steady state that does not match the GGE as the emergent double-well prevents the efficient transfer of the excitation across the chain.

\begin{figure}
\includegraphics*[width=8cm]{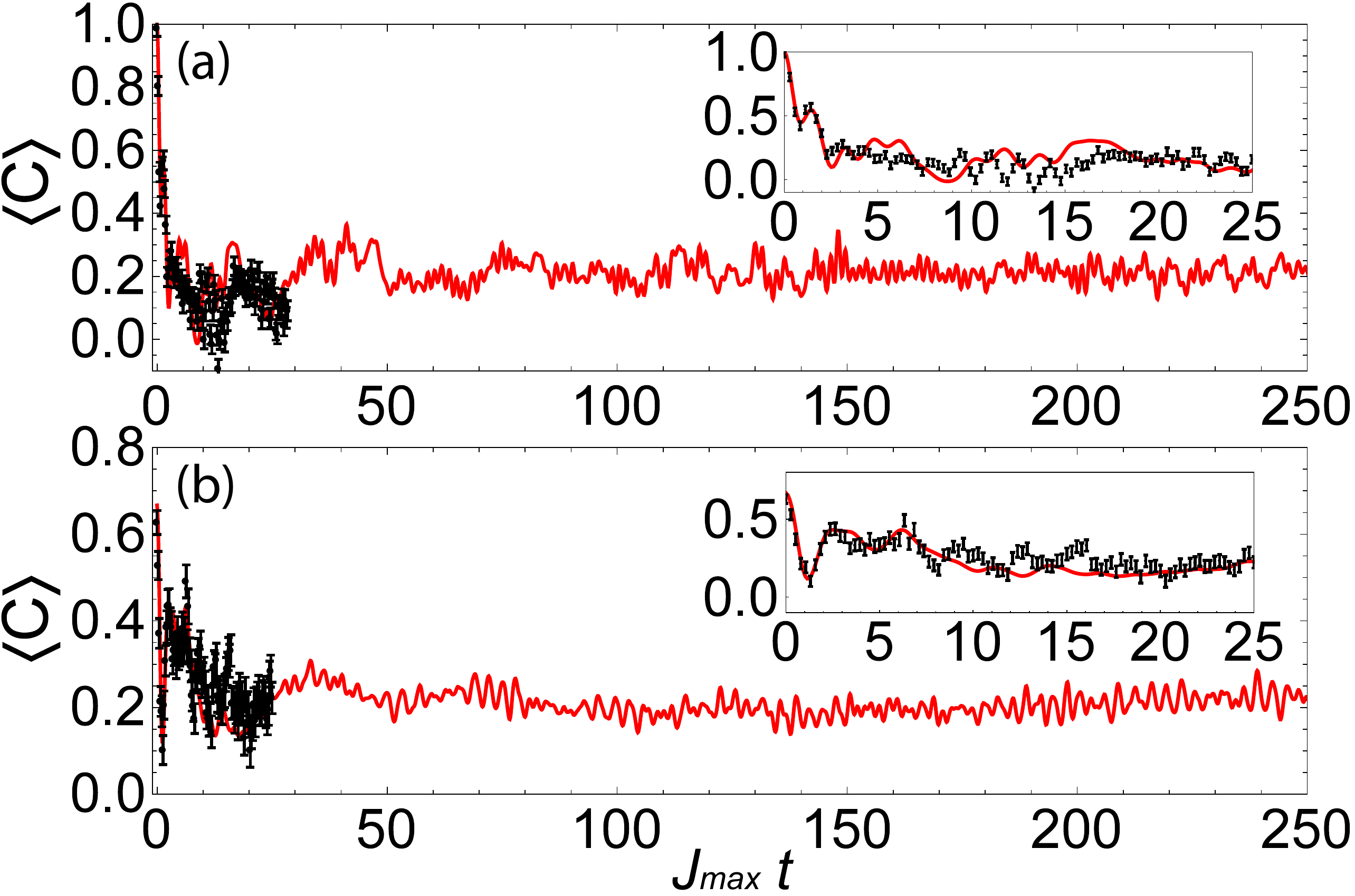}
\caption{\textbf{General feature of the model}. Comparison between a numerical simulation of the transverse field Ising model and experimental data for single (a) and double (b) initial spin excitations. The inset shows excellent agreement between experiment (black dots) and numerics (red line) accounting for experimental noise for $\langle C \rangle$. For both the single and double spin flip cases, the prethermal state persists for much longer than the experimental timescale and eventually relaxes to the thermal state.  Error bars, 1 s.d. }
\label{fig3}
\end{figure}


We show the prethermal state's robustness to weak interactions, similar to many body localization \cite{Anderson1958,Basko2007}, by preparing initial states with two spin excitations. In this case, the multiple spin flips increase the size of the integrability breaking part of the Hamiltonian which represents weak interactions between the spin-wave bosons. However, the prethermal state still persists. For better contrast between the prethermal and GGE predicted values of $\langle C \rangle$, we flip the second and fourth spin such that $\lvert\psi_{0}\rangle=\lvert\downarrow\uparrow\downarrow\uparrow\downarrow\downarrow\downarrow\rangle$. We emphasize that the observed prethermalization and departure from the GGE prediction is not sensitive to the specific choice of initial state in the thermodynamic limit. But for a small sized system, these initial states offer us the maximum signal. As before we also prepare the mirror image of the initial state by exciting the fourth and sixth ions ($\lvert\psi_{0}\rangle=\lvert\downarrow\downarrow\downarrow\uparrow\downarrow\uparrow\downarrow\rangle$). We observe relaxation to the value predicted by the GGE for short-range interactions, but with long-range interactions we see a prethermal state that strongly deviates from the GGE (bottom panel of Fig. \ref{fig2main}).

In Fig. \ref{fig3} we plot the experimental evolution of the prethermal state for both the double and single spin flip initial states along with a long-time numerical simulation under (\ref{eq:Ham}) accounting for known experimental noise (methods). The plots demonstrate excellent agreement between numerical simulations and experimental data and confirm that the prethermal states persist well beyond the current experimental time limit. Due to the non-conservation of the number of spin-wave bosons for any finite B and the interactions between them, the system will eventually relax to the thermal equilibrium in the thermodynamic limit, however relaxation to the GGE may or may not be seen depending on the range of interactions (methods). 

\begin{figure}
\includegraphics*[width=8.5cm]{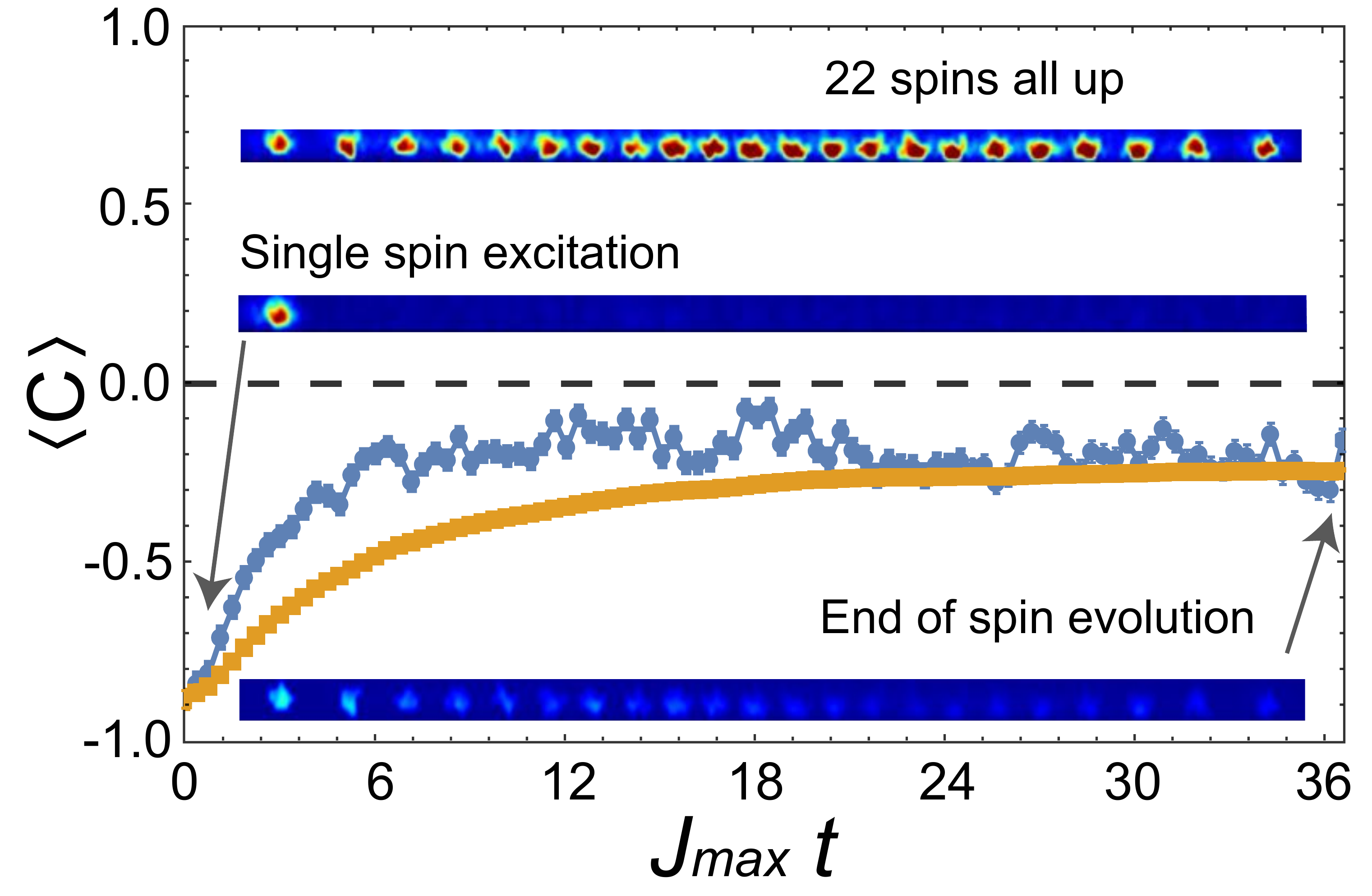}
\caption{\textbf{Scaling to larger system size}. Time evolution (light blue) and cumulative time average (orange) of $\langle C \rangle$ with false color pictures of the 22 ion chain, where the brightness of each ion is determined by the value of $\langle \overline{\sigma_i^z} \rangle$.  The ions fluoresce during detection when in the $\lvert\uparrow\rangle_{z}$ state (top picture). We initialize the spins with a single spin excitation on the left end (middle picture). After evolving for 36 $J_{\textrm{max}}$ the spin excitation is delocalized, but its average position remains on the left half of the chain (bottom picture).  Error bars, 1 s.d.}
\label{fig4}
\end{figure}

To demonstrate that the prethermal state we observe is not sensitive to system size, we repeat the experiment in a chain of 22 ions, the largest ion chain used for quantum simulation in the literature to date. The time evolution and cumulative time average of $\langle C \rangle$ are depicted in Fig. \ref{fig4}. Experimentally, as well as in the analytic result in the thermodynamic limit (methods),  we see the system relaxes to a similar quasi-equilibrium state as before that clearly has memory of the initial state.

We point out the observed prethermalization and deviation from the GGE should disappear if the system is subject to periodic boundary conditions, regardless of system size. Here we emphasize that the long-range interactions make the boundary conditions relevant for bulk properties, and thus, changing the boundary conditions can impact an extensive number of eigenstates. The effects of long-range interactions on many-body dynamics are far richer than the effect discussed in the current experiment. For sufficiently long-range interactions, the notion of locality breaks down and quasi-particles in the system can travel at divergent velocities for thermodynamic systems, potentially leading to dramatically different thermalization/prethermalization time scales in certain systems \cite{Lieb1972, Richerme2014, Jurcevic2014}. We believe that the current experiment, as well as the platform it is built upon, will pave the way to a more complete understanding of the fundamental role long-range interactions play in the quench dynamics and emergent statistical physics of quantum many-body systems. 

\section{Acknowledgements}
This work is supported by the ARO Atomic and Molecular Physics Program, the AFOSR MURI on Quantum Measurement and Verification, the IARPA LogiQ program, the NSF Physics Frontier Center at JQI, NSF QIS, AFOSR, ARL CDQI, and ARO MURI.

\bibliographystyle{prsty}
\bibliography{lotsofrefs}
 
\section{METHODS}
\renewcommand{\figurename}{Extended Data Figure}
\setcounter{figure}{0}
\subsection{Effective Hamiltonian Generation}

We generate spin-spin interactions by applying spin-dependent optical dipole forces to ions confined in a 3-layer linear Paul trap with a 4.8 MHz radial frequency. Two off-resonant laser beams with a wavevector difference $\delta k$ along a principal axis of transverse motion globally address the ions and drive stimulated Raman transitions. The two beams contain a pair of beatnote frequencies symmetrically detuned from the resonant transition at $\nu_0=12.642819$ GHz by a frequency $\mu$, comparable to the transverse motional mode frequencies. In the Lamb-Dicke regime, this results in the Ising-type Hamiltonian in Eq. (1) \cite{Molmer1999,Porras2004,Kim2009} with
\begin{equation}
J_{ij}=\frac{\hbar(\delta k)^{2}\Omega^{2}}{2M}\sum_{m=1}^{N}\frac{V_{i,m}V_{j,m}}{\mu^{2}-\omega_{m}^{2}}\label{eq:J},
\end{equation}
where $\Omega$ is the global Rabi frequency, $\hbar$ is the reduced Planck's constant, $V_{i,m}$ is the normal mode matrix \cite{James1998}, and $\omega_m$ are the transverse mode frequencies. The coupling profile may be approximated as a power-law decay $J_{ij}\approx J_{max}/|i-j|^\alpha$, where in principle $\alpha$ can be tuned between 0 and 3 by varying the laser detuning $\mu$ or the bandwidth of $\omega_m$ \cite{Porras2004,Kim2009}. For the 7 ion data in this work, $\alpha$ was tuned to 0.55 (long-range interactions) and 1.33 (short-range interactions) by changing the bandwidth of $\omega_m$. By asymmetrically adjusting the laser beatnote detuning $\mu$ about the carrier by a value of $2B$ we apply a uniform effective transverse magnetic field of ${B}\sigma_{i}^{z}$ \cite{AaronLeePhd}.

\subsection{Single spin flip initialization and site-resolved detection}

\begin{figure}
\includegraphics*[width=0.75\columnwidth]{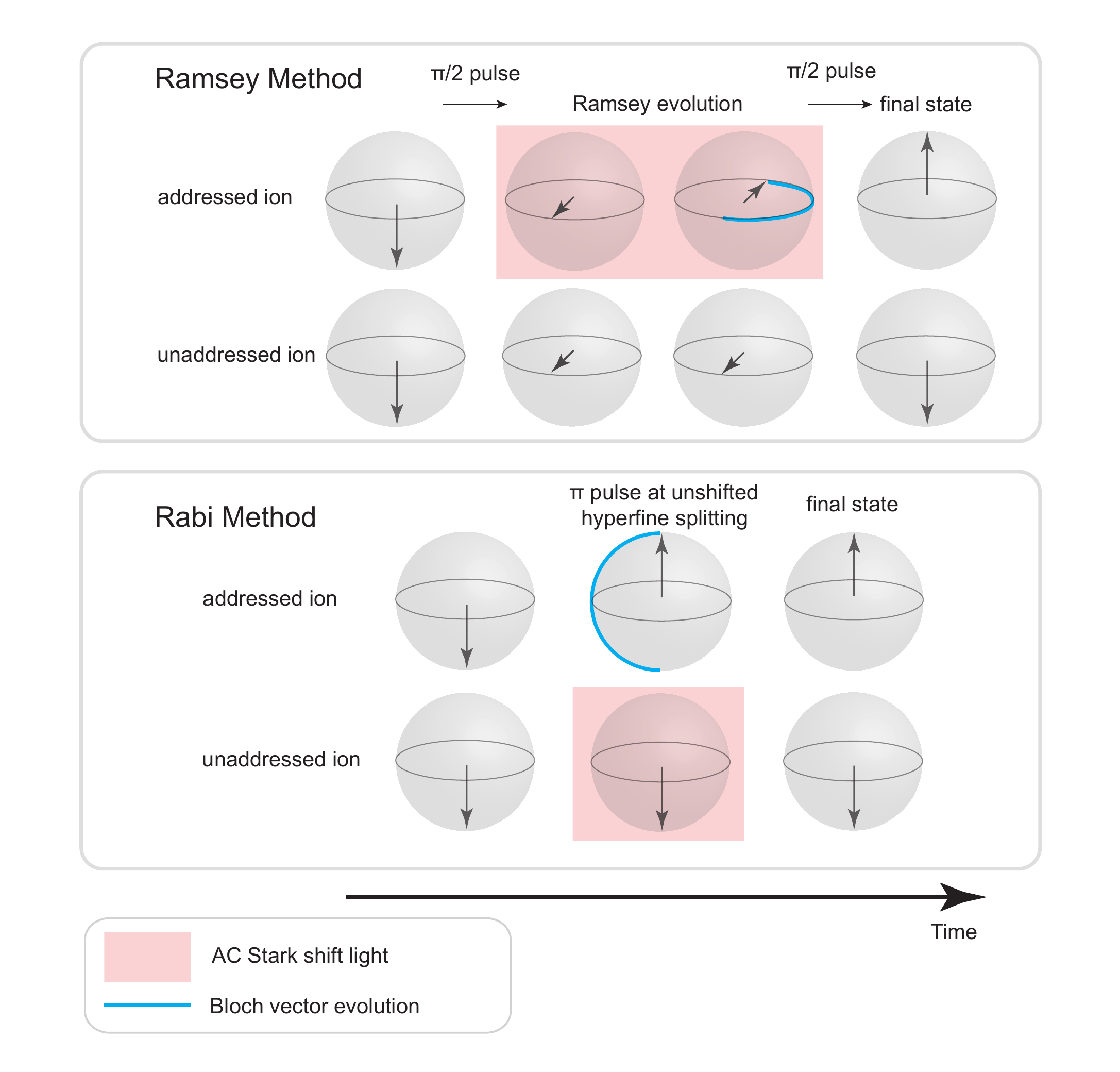}
\caption{Top box: We use a Ramsey method of preparing individual spin flips when the ion spacing is larger than the individual addressing beam waist. The spins are first optically pumped to $\ket{\downarrow}_{z}$, and then a global $\pi/2$ rotation brings them into the $xy$-plane of the Bloch sphere. The system then evolves with a Stark shift applied to the ions to be flipped until the Stark shifted spins are $\pi$ out of phase compared to the other spins. Then a global $\pi/2$ rotation brings the spins back into the $z$-basis. Bottom box: We use a Rabi method of preparing spin excitations when the ion spacing is smaller than the individual addressing beam waist. In this case the ions are optically pumped to $\ket{\downarrow}_{z}$ and then a large Stark shift is applied to all of the ions except ones to be flipped while a global $\pi$ rotation is simultaneously performed at the hyperfine splitting between the effective spin states.}
\label{stateprep}
\end{figure}

We initialize individual spin excitations using a tightly focused laser beam to imprint a fourth order AC Stark shift \cite{Lee2016} in conjunction with a Ramsey\cite{Ramsey1990} or Rabi sequence as seen in Extended Data Fig. (\ref{stateprep}). When the ion spacing is larger than the beam waist of the individual-ion addressing laser as is the case for the 7 ion short range data we employ a Ramsey method. This consists of first optically pumping the spins to $\lvert\downarrow_{z}\rangle$. Then we globally perform a $\pi/2$ rotation so that all of the spins are in $\lvert\downarrow_{x}\rangle$. Using the individual-ion addressing beam, a Stark shift is applied to the spins to be flipped, and then we allow the chain to evolve until these spins are $\pi$ out of phase compared to the spins without an applied Stark shift. Afterwards, a global $\pi/2$ rotation brings the spins back into the $z$-basis. With this method, individual spin flips can be prepared with a fidelity of $\sim0.97$, while $N$ spin flips can be achieved with a fidelity of $\sim(0.97)^{N}$.

We employ the Rabi method for the long range interacting data because site-resolved Stark shifts can no longer be applied since the ion separation is smaller than the beam waist. Here, we apply a large Stark shift to all of the spins except the ones to be flipped and apply a global $\pi$ pulse at the hyperfine splitting between the two effective spin levels. Thus, only the ions without an applied Stark shift are flipped. This approach has a single and $N$ spin flip fidelity of $\sim0.85$ and $\sim(0.85)^{N}$ respectively. 

After quenching to and allowing time evolution under our spin Hamiltonian, we measure the spin projections of each ion along the $z$ direction of the Bloch sphere. We expose the ions to a laser beam that addresses the cycling transition $^2$S$_{1/2}\ket{F=1}$ to $^2$P$_{1/2}\ket{F=0}$ for \mbox{3 ms}. Ions fluoresce only if they are in the state $\ket{\uparrow}_z$. This fluorescence is collected through an NA=0.23 objective and imaged using an intensified CCD camera with single-site resolution.

To discriminate between `bright' and `dark' states  ($\ket{\uparrow}_z$ and $\ket{\downarrow}_z$ respectively), we begin by calibrating the camera with 1000 cycles each of all-bright and all-dark states. For the bright states, the projection of the 2D CCD image onto a one-dimensional row gives a profile comprised of Gaussian distributions at each ion location. We perform fits to locate the center and fluorescence width of each ion.

We achieve single-shot discrimination of individual ion states in the experimental data by fitting the captured one-dimensional profile to a series of Gaussian distributions with calibrated widths and positions but freely-varying amplitudes. These extracted values for each ion are then compared with a threshold found via Monte-Carlo simulation to determine whether the measured state was `bright' or `dark'. Our discrimination protocol also gives an estimate of the detection error (e.g. misdiagnosing a `bright' ion as `dark'), which is typically of order $\sim 5\%$. Corrected state probabilities (along with their respective errors) are found following the method outlined in \cite{Shen2012}, which also takes into account errors due to quantum projection noise.

\subsection{Experimental noise sources and their influence on the thermalization dynamics.}

As discussed in the text, there are fluctuations on the interaction strength $J_{ij}$, which originate from noise on the laser intensity and $\omega_m$ \cite{Johnson2016}. This noise is slow compared to a single experiment, but fast compared to the thousands of experiments it takes to complete a data set.  Averaging over this classical noise leads to dynamics that resemble a running time average, because the fast temporal oscillations are effectively canceled out by the fluctuating $J_{ij}$. To account for this, the numeric simulations average over a small range of coupling strengths (standard deviation of $ 0.12 J_{max}$).

Another source of noise is the fourth order AC Stark shift from the M$\text{\o}$lmer-S$\text{\o}$rensen interaction laser sidebands. This noise also has a negligible effect, since we are in the regime of large transverse fields and this Stark shift term only adds a small global $\sigma_z$ fluctuation of about 30 Hz, on top of the 10 kHz transverse field applied. We experimentally verify this by applying a global offset field of $\pm$ 400 Hz and observe no difference in the observed prethermalization. The relaxation dynamics are robust against these experimental noise sources, however it is sensitive to asymmetries of the spin-spin coupling matrix.  

\subsection{Measuring the spin-spin coupling matrix}

\begin{figure}
\includegraphics*[width=\columnwidth]{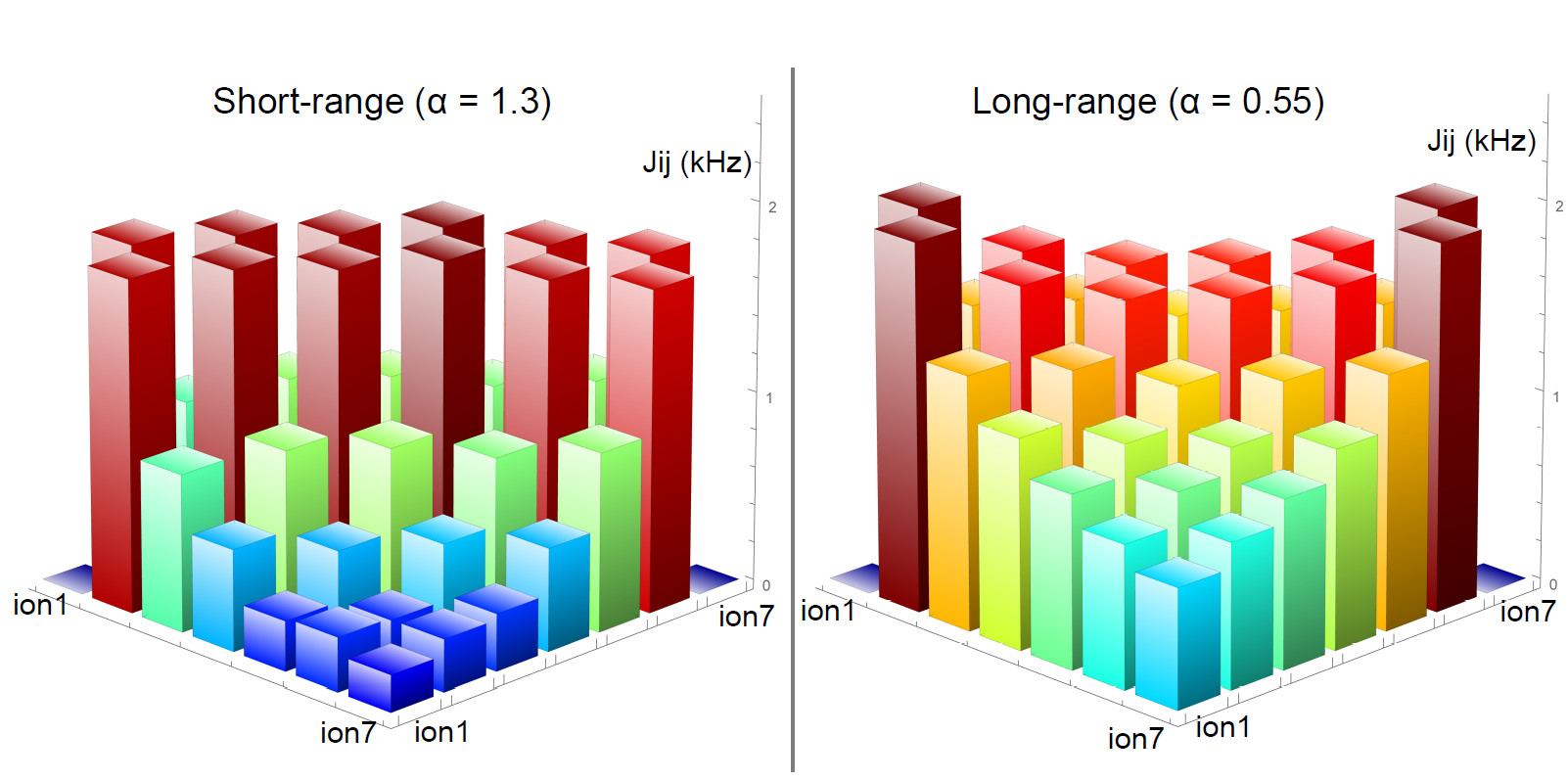}
\caption{We directly measure the spin-spin coupling matrix with 7 ions for both short (left matrix) and long (right matrix) range interactions and see it is symmetric. To measure the coupling between two spins we transfer all except the spins of interest out of the interaction space. This is done by shelving the other spins in one of the Zeeman levels.}
\label{fig2}
\end{figure}

We directly measure the spin-spin coupling matrices for seven ions for both long and short range interactions and ensure it is symmetric as seen in Extended Data Fig. \ref{fig2}. In order to measure the strength of the interaction between two spins, we shelve all but the two ions of interest out of the interaction space and directly observe their time evolution. This is done by first performing individual rotations on these two ions to the $\ket{\uparrow}_z$ as outlined above. Then we perform a global $\pi$ rotation between $\ket{\downarrow}_z$, $^2$S$_{1/2}\ket{F=0,m_F=0}$, and one of the Zeeman states, $^2$S$_{1/2}\ket{F=1,m_F=-1}$, which takes the other 5 spins out of the interaction space. We then apply the Hamiltonian which now only acts on the two remaining spins.

\subsection{Justification for postselection}
As noted above the initial $N$ spin flip fidelity is approximately $(0.97)^{N}$ and $(0.85)^{N}$ for short and long range interactions respectively. Numerically, we find that the number of spin excitations is essentially constant on the experimental timescale which is expected because the transverse field Ising model can be mapped to an XY model for sufficiently large transverse field and the XY model conserves the number of spin excitations \cite{Richerme2014}. However, experimentally we observe significant leakage out of the $N$ excitation subspace with less than $50\%$ remaining at the end of the evolution. Thus, we postselect the data for the correct number of spin excitations to eliminate the effects of imperfect state preparation, detection error, and small deviations from our model Hamiltonian due to unwanted excitations of the phonon modes.

\subsection{The spin-boson mapping and the generalized Gibbs ensemble}
To explain our observed prethermalization, it is convenient to
map the spins into bosons by using the Holstein-Primakoff transformation: $\sigma_{i}^{z}=2a_{i}^{\dagger}a_{i}-1$,
$\sigma_{i}^{+}=a_{i}^{\dagger}\sqrt{1-a_{i}^{\dagger}a_{i}}$ . We
will assume that the average spin excitation density $\bar{n}=\sum_{i}\langle a_{i}^{\dagger}a_{i}\rangle/N$
is much smaller than $1$. This assumption is justified in our experiments
because our initial states have small spin excitation densities
and we set $\max(J_{ij})\ll B$ so the amount of $\bar{n}$ that
will be dynamically created is small {[}$\sim(\max(J_{ij})/B)^{2}${]}.
Therefore to the lowest order we can approximate $\sigma_{i}^{+}\approx a_{i}^{\dagger}$,
and (1) reduces to an integrable Hamiltonian $H_{0}$ made of
non-interacting bosons

\begin{align}
H_{0} & =\sum_{i<j}J_{ij}(a_{i}^{\dagger}a_{j}+a_{i}^{\dagger}a_{j}^{\dagger}+h.c.)+2B\sum_{i}a_{i}^{\dagger}a_{i},\label{eq:H0def}\\
H_{1} & =H-H_{0}.
\end{align}
Here $H_{1}$ contains interactions between the bosons which are parametrically
small in $\bar{n}$, and, as a result, we can treat $H_{1}$ as a perturbation to
$H_{0}$. Thus, it is natural to expect the system (in the thermodynamic
limit) to first relax to a prethermal state described by the GGE of
$H_{0}$, and to later relax to a thermal state described by the full
$H$. Naively, we expect the thermalization to happen at a time scale
much longer than the relaxation to GGE, based on the different energy
scales of $H_{0}$ and $H_{1}$. However, this is not always the case,
as discussed below.

To explicitly define the GGE of $H_{0}$, we would need to first diagonalize
$H_{0}$ and find the integrals of motion. Note that (\ref{eq:H0def})
only involves $J_{ij}$ for $i<j$. For convenience, we will define
a new matrix $\mathcal{J}$ such that $\mathcal{J}_{ii}=0$ and $\mathcal{J}_{ij}=\mathcal{J}_{ji}=J_{ij}$
for $i<j$. $H_{0}$ can be rewritten as
\begin{equation}
H_{0}=\sum_{i,j}\mathcal{J}_{ij}\left[a_{i}^{\dagger}a_{j}+\frac{1}{2}(a_{i}^{\dagger}a_{j}^{\dagger}+a_{i}a_{j})\right]+2B\sum_{i}a_{i}^{\dagger}a_{i}.
\end{equation}

An orthogonal matrix $\mathcal{V}$ can be used to diagonalize the
matrix $\mathcal{J}$ as $\sum_{i,j}\mathcal{V}_{ik}\mathcal{J}_{ij}\mathcal{V}_{jk^{\prime}}=\nu_{k}\delta_{kk^{\prime}}$,
where $\{\nu_{k}\}$ are the eigenvalues of matrix $\mathcal{J}$.
Introducing $c_{k}=\sum_{i}\mathcal{V}_{ik}a_{i}$, we have
\begin{equation}
H_{0}=\sum_{k}\left[(\nu_{k}+2B)c_{k}^{\dagger}c_{k}+\frac{1}{2}\nu_{k}(c_{k}^{\dagger}c_{k}^{\dagger}+c_{k}c_{k})\right].
\end{equation}

Next, we perform a Bogoliubov transformation $c_{k}=\cosh(\theta_{k})d_{k}-\sinh(\theta_{k})d_{k}^{\dagger}$
with $\theta_{k}=\frac{1}{2}\tanh^{-1}(\frac{\nu_{k}}{\nu_{k}+2B})$
to fully diagonalize $H_{0}$:
\begin{equation}
H_{0}=\sum_{k}\epsilon_{k}d_{k}^{\dagger}d_{k},\hspace{1em}\epsilon_{k}\equiv2\sqrt{B(B+\nu_{k})}\label{eq:H0}.
\end{equation}

Since integrable models have an extensive number of conserved quantities that are not taken into account by canonical ensembles from statistical mechanics the GGE was developed to make predictions about equilibrium values of observables in these systems by incorporating the additional integrals of motion \cite{Kollar2011,Jaynes1957,Rigol2007,Cazalilla2006,Caux2012,Ilievski2015}. The GGE for $H_{0}$ is defined as
\begin{equation}
\rho_{GGE}=\frac{e^{-\sum_{k}\lambda_{k}d_{k}^{\dagger}d_{k}}}{\text{Tr}(e^{-\sum_{k}\lambda_{k}d_{k}^{\dagger}d_{k}})},
\end{equation}
with $\lambda_{k}$s determined by $\langle d_{k}^{\dagger}d_{k}\rangle_{0}=\langle d_{k}^{\dagger}d_{k}\rangle_{GGE}$,
where the notation $\langle\cdots\rangle_{0}$ denotes the expectation
value in the initial state $|\psi_{0}\rangle$, and the notation $\langle\cdots\rangle_{GGE}$
denotes the expectation value in $\rho_{GGE}$. We observe relaxation to the GGE if for any local observable $\hat{O}$, $\langle O(t)\rangle\approx\text{tr}(O\rho_{GGE})$. The GGE is equivalent to the grand canonical  ensemble for a general quantum system when the only conserved quantities are particle number and total energy\cite{Rigol2007}.

Using the fact that
our initial state $|\psi_{0}\rangle$ is always a Fock state in the
basis of $\{a_{i}^{\dagger}a_{i}\}$, the value of $\langle d_{k}^{\dagger}d_{k}\rangle_{0}$
can be calculated by the following formula
\begin{align}
\langle d_{k}^{\dagger}d_{k}\rangle_{0} & =\cosh(2\theta_{k})\sum_{i}\mathcal{V}_{ik}^{2}\langle a_{i}^{\dagger}a_{i}\rangle_{0}+\sinh^{2}(\theta_{k}).
\end{align}

To calculate the expectation values of $\sigma_{i}^{z}=2a_{i}^{\dagger}a_{i}-1$
in the GGE, we use the following equations 
\begin{align}
\begin{split}
\langle a_{i}^{\dagger}a_{i}\rangle_{GGE} & =\langle\sum_{k,k^{\prime}}\mathcal{V}_{ik}\mathcal{V}_{ik^{\prime}}c_{k}^{\dagger}c_{k^{\prime}}\rangle_{GGE}\\
 & =\sum_{k}\mathcal{V}_{ik}^{2}\langle\cosh(2\theta_{k})d_{k}^{\dagger}d_{k}+\sinh^{2}(\theta_{k})\rangle_{GGE}\\
 & =\sum_{k}\mathcal{V}_{ik}^{2}[\cosh(2\theta_{k})\langle d_{k}^{\dagger}d_{k}\rangle_{0}+\sinh^{2}(\theta_{k})],
\end{split}
\end{align}
where we use the fact that $\rho_{GGE}$ is diagonal in the Fock basis
of $\{d_{k}^{\dagger}d_{k}\}$, so $\langle d_{k}^{\dagger}d_{k^{\prime}}\rangle_{GGE}=\langle d_{k}^{\dagger}d_{k^{\prime}}^{\dagger}\rangle_{GGE}=0$
for $k\ne k^{\prime}$. 

\subsection{Single-particle properties of $H_{0}$}

Since $v_{k}/B$ is small, we can expand $\epsilon_{k}$ and $d_{k}$
in $\nu_{k}/B$ to the leading order:
\begin{equation}
\epsilon_{k}\approx2B+\nu_{k},\hspace{1em}d_{k}\approx\sum_{i}\mathcal{V}_{ik}(a_{i}+\frac{\nu_{k}}{4B}a_{i}^{\dagger}).\label{eq:bogospectrum}
\end{equation}
This means to understand the single-particle properties of $H_{0}$
(Eq.\,\ref{eq:H0}), we just need to understand the properties of
the eigenvalues $\{\nu_{k}\}$ and eigenvectors $\mathcal{V}$ of
the matrix $\mathcal{J}$. We emphasize that the matrix $J$ defined in (\ref{eq:J}) differs from the matrix $\mathcal{J}$
used in defining $H_{0}$, because $J_{ii}\ne0$ by the above definition
(\,\ref{eq:J}). This is because $J_{ii}$ has no physical consequence
in the Ising Hamiltonian as $(\sigma_{i}^{x})^{2}=1$.

\subsection{Eigenvalues and eigenvectors of $J$}

Let us first try to understand the properties of the eigenvalues and
eigenvectors of the matrix $J$. To make this possible, we will need
to approximate the spacing between ions to be uniform. While this
is not true in the current experiment due to the harmonic trapping potential,
the inhomogeneity in ion spacing is not responsible for the observed
prethermalization \cite{Gong2013} and from now on
we will assume that the ions are equally spaced.

We now write down the motional Hamiltonian of $N$ ions trapped along
the $z$ direction ignoring the ions' motion along the $y$ direction
for simplicity since we barely couple the spins to the phonons in that direction:

\begin{eqnarray}
H_{m} =& \sum_{i=1}^{N}\left[\frac{p_{i,x}^{2}}{2M}+\frac{p_{i,z}^{2}}{2M}+V(z_{i})+\frac{1}{2}M\omega_{x}^{2}x_{i}^{2}\right]\\
&+\frac{Q^{2}}{4\pi\epsilon_{0}}\sum_{i=1}^{N}\sum_{j=1}^{i-1}\frac{1}{\sqrt{(z_{i}-z_{j})^{2}+(x_{i}-x_{j})^{2}}}.\label{Hm}
\end{eqnarray}
Here $\{x_{i},z_{i},p_{i,x},p_{i,z}\}$ are respectively the coordinates
and momenta of the $i^{th}$ ion in the $x$ and $z$ directions.
$M$ and $Q$ are the mass and charge of each ion, and $\omega_{x}$
is the transverse trapping frequency. The ions will be equally spaced
with a spacing $a_{0}$ if $V(z)=-\frac{Q^{2}}{4\pi\epsilon_{0}a_{0}}\log(1-z^{2}/L^{2})$,
with $L=Na_{0}/2$.

By expanding the Coulomb interaction around the ions' equilibrium positions
up to second order in position, the motional Hamiltonian in
the $x$ direction can be written as:
\begin{equation}
H_{mx}=\sum_{i=1}^{N}\frac{p_{i,x}^{2}}{2M}+\frac{1}{2}M\left(\sum_{i=1}^{N}\omega_{x}^{2}x_{i}^{2}-\omega_{z}^{2}\sum_{i,j=1}^{N}K_{ij}x_{i}x_{j}\right),
\end{equation}
where we have set $\omega_{z}\equiv\sqrt{\frac{Q^{2}}{4\pi\epsilon_{0}Ma_{0}^{3}}}$
as an ``effective'' axial trapping frequency. The dimensionless
matrix $K$ characterizes the \emph{dipolar interactions} between
ions:
\begin{equation}
K_{i\ne j}=-|i-j|{}^{-3},\hspace{1em}K_{ii}=-\sum_{j\ne i}K_{ij}\label{eq:K},
\end{equation}

The exact analytical expressions for the eigenvalues $\{\kappa_{m}\}$
and eigenvectors $\{V_{i,m}\}$ of $K$ cannot be obtained. But we
can employ a first-order perturbation theory and assume that the eigenvectors
of $K$ are approximately the same as those of a nearest-neighbor
coupling matrix $K$. As a result:
\begin{eqnarray}
V_{i,m} & \approx & \begin{cases}
\sqrt{1/N}, & m=0,\\
\sqrt{\frac{2}{N}}\cos[\frac{m\pi}{N}(i-\frac{1}{2})], & m=1,2,\cdots N-1,
\end{cases}\label{eq:bs}\\
\kappa_{m} & \approx & \sum_{r=1}^{N/2}\frac{2-2\cos\frac{mr\pi}{N}}{r^{3}},\hspace{1em}(m=0,1,\cdots N-1)\label{eq:kappa}.
\end{eqnarray}
Note that the $(i-\frac{1}{2})$ above ensures that the phonon modes
are either symmetric or antisymmetric under the spatial inversion
of the chain ($i\rightarrow N+1-i$).

As a result, the eigenvectors of the matrix $J$ are given by $\{V_{i,m}\}$,
and the eigenvalues are given by 
\begin{equation}
\lambda_{m}=\frac{\hbar(\delta k)^{2}\Omega^{2}}{2M(\mu^{2}-\omega_{x}^{2}+\omega_{z}^{2}\kappa_{m})}\label{lambda},
\end{equation}

Finally, we point out that importantly, $J_{ii}$ is in general
non-uniform. This can be seen in the following two limits:
\begin{enumerate}
\item When $\mu^{2}-\omega_{x}^{2}\gg\omega_{z}^{2}\kappa_{m}$ for all
$m$, we expect $J_{ij}$ to decay as $1/|i-j|^{3}$ ($\alpha\rightarrow3$
limit), and 
\begin{equation}
J_{ii}\approx\frac{\hbar(\delta k)^{2}\Omega^{2}\omega_{z}^{2}}{2M(\mu^{2}-\omega_{x}^{2})^{2}}K_{ii}.
\end{equation}
$J_{ii}$ in this limit is very close to uniform in the large $N$
limit, except for $i$ close to $1$ and $N$ (see Fig.\,\ref{fig1}a).
\item When $\mu^{2}-\omega_{x}^{2}\ll\omega_{z}^{2}\kappa_{m}$ for all
$m\ne0$ ($\alpha\rightarrow0$ limit), we can separate out the $m=0$
term and approximate $J_{ii}$ by
\begin{align}
J_{ii} & \approx\frac{\hbar(\delta k)^{2}\Omega^{2}}{2M\omega_{z}^{2}}\frac{1}{N}\left\{ 1+2\sum_{m=1}^{N-1}\frac{\cos^{2}[\frac{m\pi}{N}(i-\frac{1}{2})]}{\kappa_{m}}\right\} .
\end{align}
Note that even in the large $N$ limit, $J_{ii}$ is  non-uniform across
the entire ion chain (see Fig.\,\ref{fig1}a). An analytical formula
can be obtained if we approximate $\kappa_{m}$ by including only
the $r=1$ (nearest-neighbor) term in Eq.\,\ref{eq:kappa}, leading
to $J_{ii}\approx\frac{\hbar(\delta k)^{2}\Omega^{2}}{2M\omega_{z}^{2}}\frac{1}{N}(i-\frac{N+1}{2})^{2}+\text{constant}$.
\end{enumerate}

\subsection{Eigenvalues and eigenvectors of $\mathcal{J}$}

As shown above, when the interactions described by $J_{ij}$ are very
long-ranged, $J_{ii}$ can be rather non-uniform. This will result
in a qualitatively different structure between the eigenvalues and
eigenvectors of the two matrices $J$ and $\mathcal{J}$. To see this,
we first notice that the eigenvalues and eigenvectors of $J$ are
similar to those of the Hamiltonian for a free-particle in an square well potential. This connection
can be formalized by going into the continuum limit and introducing
a continuous momentum $q\equiv m\pi/N\in(0,\pi)$. The eigenspectrum
$\lambda(q)$ of $J$ is minimized at $q=\pi$. Expanding $\lambda(q)$
around $q=\pi$ and using Eq.\,\ref{eq:kappa}-\ref{lambda}, we
obtain
\begin{align}
\lambda(q) & \approx O((q-\pi)^{0})+\frac{\hbar(\delta k)^{2}}{2M_{\text{eff}}}(q-\pi)^{2}+O[(q-\pi)^{4}],\\
M_{\text{eff}} & \equiv M\frac{\left[\mu^{2}-\omega_{x}^{2}+4\zeta(3)\right]^{2}}{\omega_{z}^{2}\Omega^{2}\ln2},
\end{align}
corresponding to the dispersion relation of a massive particle with
an effective mass $M_{eff}$ and an effective momentum $(\delta k)q$.

The Schr$\ddot{\text{o}}$dinger equation for the above particle can be written in
the position space parameterized by a continuous coordinate $z\in[1,N]$
that replaces the discrete ion index $i\in\{1,2,\cdots N\}$:
\begin{equation}
-\frac{\hbar^{2}(\delta k)^{2}}{2M_{\text{eff}}}\frac{\partial^{2}}{\partial z^{2}}\psi(z)=E\psi(z),\label{eq:Scheq}
\end{equation}
with the boundary condition $\psi(z\le1)=\psi(z\ge N)=0$ corresponding
to that of a particle in a box potential. Here the eigenwavefunction
$\psi_{m}(z)\approx(-1)^{i}V_{i,m}$ (for $z=i$), and the eigenenergy
$E_{m}\approx\hbar\lambda_{m}$ (up to a constant shift) near $q=\pi$.

\begin{figure}
\includegraphics[width=0.49\textwidth]{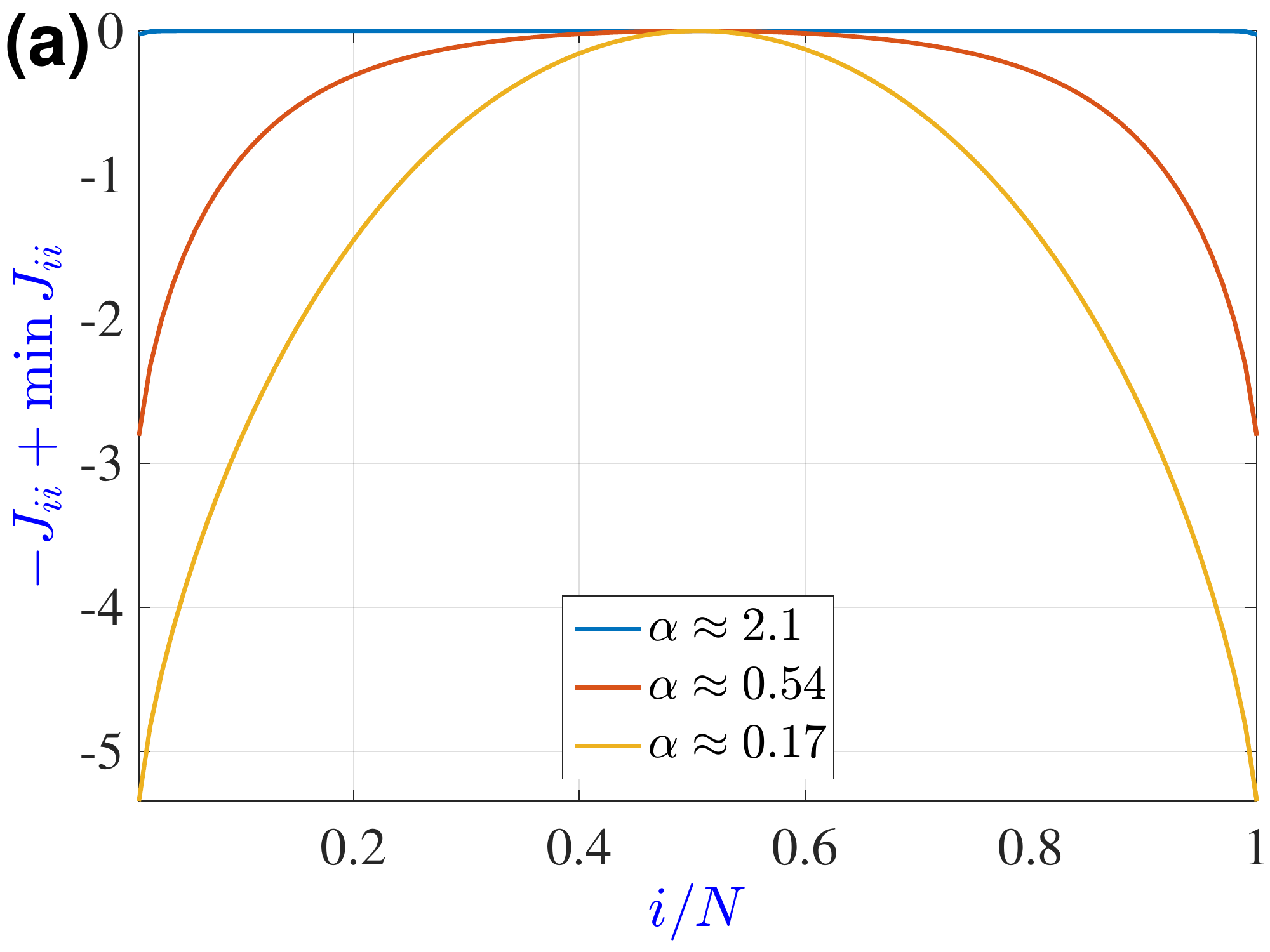}\hfill{}\includegraphics[width=0.49\textwidth]{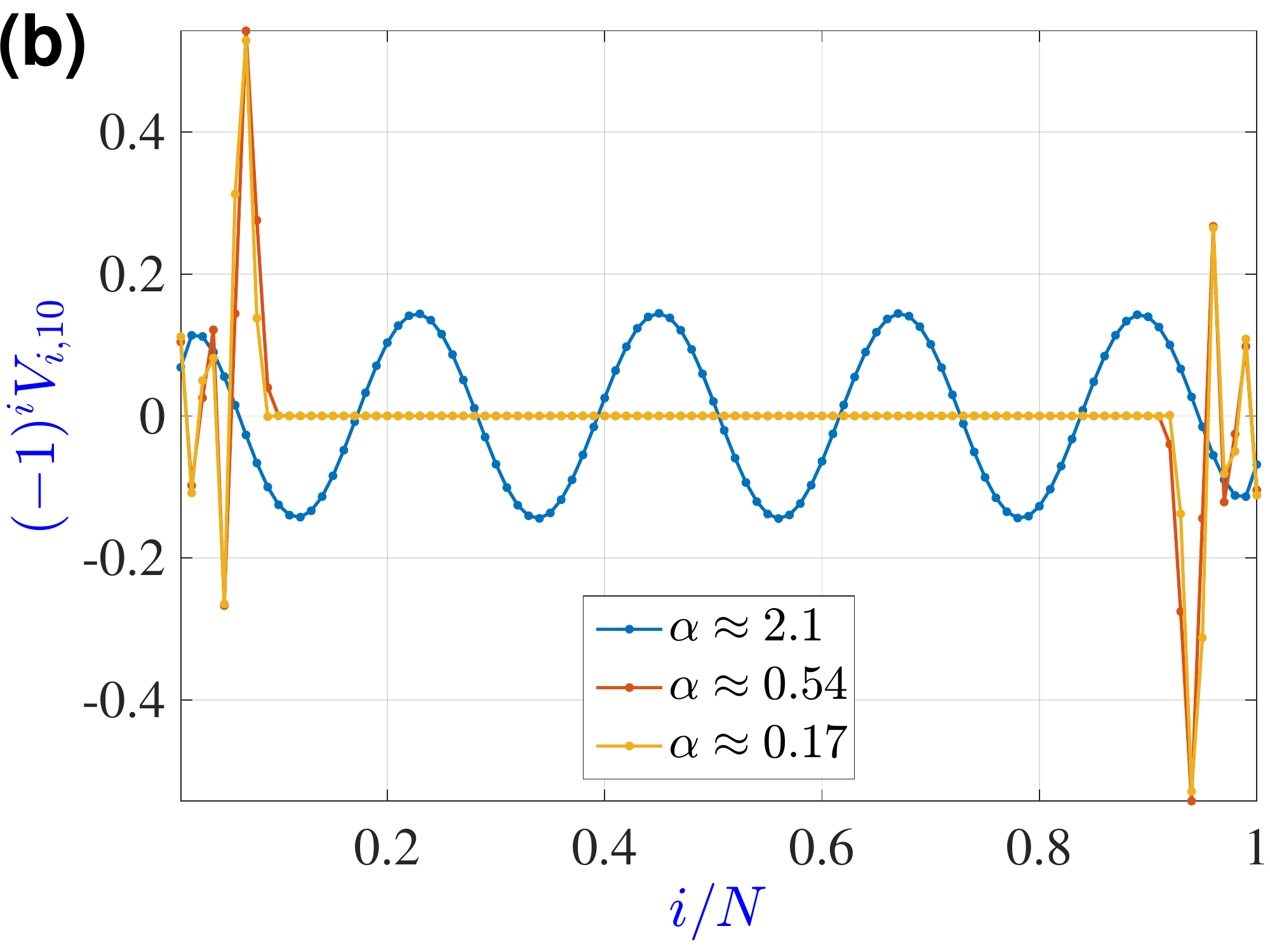}

\caption{(a) The diagonal matrix element $-J_{ii}$ {[}in arbitrary units and
shifted by $\min(J_{ii})${]} that determines the single-particle
potential $U(z)$ for $N=100$ ions. We choose the parameters that
make $J_{ij}$ decay approximately as $1/r^{\alpha}$ with different
values of $\alpha$ shown in the plot. As $\alpha$ decreases, the
potential changes continuously from nearly flat to an approximately
harmonic anti-trap. Together with two hard wall potentials at $i/N=0,1$,
the potential looks like a double well that becomes deeper for smaller
$\alpha$. (b) The eigenvector $V_{i,10}$ corresponding to the $10^{th}$
lowest eigenvalue of $\mathcal{J}$ for a $N=100$ ion chain. For
small $\alpha$'s, the eigenvector, as well as the wavefunction $\Psi(z)$
of Eq.\,(\ref{eq:Scheq-1}), is localized inside the two wells. For
large $\alpha$'s, the eigenvector is delocalized and similar to that
of a particle in a box.\label{fig1}}
 
\end{figure}

We can similarly map the eigenvalue equation of $\mathcal{J}$ to
a Schr$\ddot{\text{o}}$dinger equation of a massive particle:
\begin{equation}
-\frac{\hbar^{2}(\delta k)^{2}}{2M_{\text{eff}}}\frac{\partial^{2}}{\partial z^{2}}\Psi(z)+U(z)=\mathcal{E}\Psi(z).\label{eq:Scheq-1}
\end{equation}
 However, we now have an effective potential $U(z)$ due to the fact
that $\mathcal{J}_{ii}=0\ne J_{ii}$, and up to a constant energy
shift
\begin{equation}
U(z)\equiv\begin{cases}
\infty & z<1\text{ or }z>N\\
-J_{ii} & z=i\in\{1,2,\cdots,N\}
\end{cases}.
\end{equation}
As discussed in the previous section, for $\mu^{2}-\omega_{x}^{2}\gg\omega_{z}^{2}\kappa_{m}$
(where $\alpha\rightarrow3$), the potential $U(z)$ will be nearly
flat, and the eigenvalues and eigenvectors of $\mathcal{J}$ are similar
to those of a particle in a box. However, for $\mu^{2}-\omega_{x}^{2}\ll\omega_{z}^{2}\kappa_{m}$
(where $\alpha\rightarrow0$), $U(z)$ has the shape of a double well
potential (Extended Data Fig.\,\ref{fig1}a).

\subsection{Discussion}

Ignoring tunneling between the two deep wells of a double-well-shaped potential, the low-energy eigenstates of a massive particle in such a potential are localized orbitals inside either well (Extended Data Fig.\,\ref{fig1}b).
The tunneling rate (which is exponentially small in the height of
the barrier) splits the degeneracy of the localized orbitals in each
well and leads to pairs of symmetric and antisymmetric (upon the spatial
inversion of the chain) wavefunctions. This physical picture explains
the observed prethermalization: Initial excitations placed in the
left half of the chain will be localized for an extended period of time
under the evolution of $H_{0}$, until the tunneling between the two wells eventually
delocalizes the excitations.

Importantly, the double-well-shaped potential is emergent in our
system, because the non-interacting Hamiltonian $H_{0}$ does not contain
an inhomogeneous potential. This emergent inhomogeneity is somewhat
a surprising effect because the motional Hamiltonian (Eq.\,\ref{Hm})
and spin-motion couplings induced by the Raman lasers are all homogeneous.
The key reason is that the long-range interactions break the translational invariance of the Ising Hamiltonian
(even in the thermodynamic limit). This is in contrast to an open-boundary spin chain with short-range interactions, where it is
safe to assume translational invariance for sufficiently large system
sizes. 

The notion of a boundary starts to break down for sufficiently long-ranged
interactions, and therefore we cannot attribute the observed prethermalization
to boundary effects. Usually, boundary effects only affect a finite
number of eigenstates and do not affect local quenches in the bulk.
However, here there is extensive number of eigenstates that are localized
in the two wells described by the potential $U(z)$, and excitations
placed an extensive number of lattice sites away from the edges
are still subject to qualitatively similar dynamics.

Finally, we point out that interactions in $H_{1}$ can also delocalize
the initial spin excitations placed in one of the wells, and eventually
thermalize the system. As a result, there is an interesting interplay
between the timescales of prethermalization to the GGE and
of thermalization. If the interactions in $H_{1}$  are sufficiently
weaker than the kinetic tunneling rate in $H_{0}$, which can be achieved
by increasing the magnetic field strength $B$ or changing the
range of interactions, then we expect the system to have two prethermal
phases before thermalization, with the observed prethermalization followed by prethermalization
to the GGE of $H_{0}$. If instead the tunneling rate is sufficiently smaller
than the interactions in $H_{1}$, then the prethermal states
described by the GGE of $H_{0}$ may never appear during the time
evolution. These interesting multi-stage relaxation
processes will require future experimental investigations with longer
coherence times and larger spin chains.

\end{document}